\documentclass{epl}
\usepackage{latexsym}

\title{Joint reality and Bell inequalities for consecutive measurements}

\shorttitle{Joint realism and Bell inequalities}

\author{R. Lapiedra\thanks{E-mail address: ramon.lapiedra@uv.es (R. Lapiedra). Telephone: 96 3543077}}

\institute{Department of Astronomy and Astrophysics, University of
Valencia, 46100 Burjassot (Valencia), Spain}


\begin{document}

\maketitle

\begin{abstract}
Some new Bell inequalities for consecutive measurements are
deduced under joint realism assumption, using some perfect
correlation property. No locality condition is needed. When the
measured system is a macroscopic system, joint realism assumption
substitutes the non-invasive measurability hypothesis
advantageously, provided that the system satisfies the perfect
correlation property. The new inequalities are violated
quantically. This violation can be expected to be more severe than
in the case of precedent temporal Bell inequalities. Some
microscopic and mesoscopic situations, in which the new
inequalities could be tested, are roughly considered.
\end{abstract}

\section{1. Introduction}

Besides the ordinary Bell inequalities [1, 2] for entangled
systems, there also exist so-called temporal Bell inequalities
[3-5] for a single system. In the seminal paper of Leggett and
Garg [6], the authors consider a macroscopic system and make two
general assumptions:

i) Macroscopic realism: ``A macroscopic system with two or more
macroscopically distinct states available to it will at all times be in one
or the other of these states''.

ii) Noninvasive measurability (NIM): ``It is possible, in principle, to
determine the state of the system with arbitrarily small perturbation in its
subsequent dynamics''.

With these two assumptions, these authors prove some temporal Bell
inequalities for such a macroscopic system, where the measurement
times, \textit{t}$_{i}$, play the role of the polarizer settings
in the ordinary Bell inequalities\textbf{. }NIM assumption is
obviously not valid for quantum systems, and has been criticized
for macroscopic systems [7]. In spite of these criticisms, it
seems that the idea of an \textit{ideal negative experiment}, or
alternatively the \textit{coupling of the system to a microscopic
probe}, as explained in [6], can change NIM into a reasonable
assumption.

Whatever it be, the main purpose of the present paper is to prove
some new Bell inequalities for consecutive measurements, retaining
the realism of assumption i), but changing the NIM for a new
assumption, that encompass the above assumption i), and becomes
extremely natural and plausible if NIM is assumed, but not
necessarily the reverse way. We will call this new assumption the
\textit{joint reality} assumption and we will state it below.
Contrarily to what happens in the temporal Bell inequalities of
Leggett and Garg, we here deal again with the original polarizer
settings, instead of the above measurement times. Thus, in our
Bell inequalities, we will perform different kinds of measurements
in \textit{a time ordered way}, on a non extended system.

The new Bell inequalities for consecutive measurements, which
apply to any macroscopic or microscopic system, will be valid
provided that the above assumption holds and the system obeys a so
called perfect correlation property. This property is always valid
for quantum systems and could be valid for macroscopic systems. In
any case, for these last systems, one can always test whether the
property is actually satisfied or not.(This perfect correlation
property will be properly described below. Then, it will become
clear that its validity is a first indication of the possible
correctness of the original NIM assumption). The new Bell
inequalities will be violated by quantum mechanics.

We will now go on to state this \textit{joint reality} assumption,
which will substitute the above two assumptions i) and ii).
Consider an ensemble of systems $S$, prepared in some way at an
initial time. $S$ can be either macroscopic or not, and has a
dichotomic magnitude, $M$, that is, a magnitude which only takes
two values, say $\pm 1$. We will measure $M$ for three different
values, $a$, $b$, and $c$, of an external parameter. (An example
could be a one half spin particle measured on three different
directions). Then, the above joint reality assumption assumes the
joint existence of a reality behind any obtainable measurement
outcome. More precisely we will denote this joint reality by
$(a^{\alpha }b^{\beta }c^{\gamma })$, where $\alpha $, $\beta $,
$\gamma =\pm $. That is, $(a^{\alpha }b^{\beta }c^{\gamma })$ is
the reality such that, if we took a measurement above, for the
parameter value $a$, we would obtain $\alpha 1$, i.e., $+1$ or
$-1$, according to the value of $\alpha $. Similar for the other
\textit{directions} $b$ and $c$. Notice that, like d'Espagnat in
Ref. [8], and Wigner in Ref. [9], we assume the existence of a
reality for all the possible results of all possible measurements,
even if each actual measurement is taken in a single randomly
chosen direction. As we have commented above, this kind of reality
is a very natural assumption as far as one assumes NIM. We will
comment below why it will be also interesting to assume initially
this kind of reality in the case of a quantum system, where
obviously the NIM assumption is non valid.

On the other hand, suppose we perform two immediately consecutive
measurements for the same external parameter value. Then, we
assume that the corresponding outcomes are \textit{perfectly
correlated}, i. e., if the first measurement value is $+1$, the
second one is always $+1$, and likewise for a value of $-1$. This
will be called the \textit{perfect correlation }property.
Obviously, this property is always satisfied when $S$ is a quantum
system, a \textit{qubit }in this paper\textit{ }(i.e., a quantum
system whose space of states is 2-dimensional), and we take pairs
of immediately consecutive measurements for different
\textit{directions} randomly chosen among the same three different
external parameter values. For a macroscopic system, perfect
correlation property can be expected to be valid as far as the NIM
assumption is correct. But, here, the question is simply whether
experience will show or not that the property is satisfied. If it
does, we meet the conditions to prove our inequalities. Otherwise,
we could not prove them.

Thus, under joint realism assumption, using the perfect
correlation property, we will prove some new Bell inequalities
where the measurement times, $t_{i}$, of Ref. [6] will be replaced
by the above parameter values, $a$, $b$, and $c$. In Ref. [10] a
similar problem is addressed. The authors prove the inequalities
for consecutive measurements which are analogous to the ordinary
CHSH inequalities [2], under the ``locality in time'' assumption.
We will not use here this doubtful assumption, which will be
replaced by the above joint reality assumption and the use of the
perfect correlation property.

\section{2. Proving the new Bell inequalities for consecutive measurements}

Let us consider the above perfectly correlated system, $S$, with
its dichotomic magnitude, $M$, measured randomly and consecutively
for the external parameter values $a$, $b$, and $c$. We want to
prove some Bell inequality for the outcomes of these measurements,
assuming joint realism and using perfect correlation. In order to
do so, we will adapt to our case a proof of an ordinary Bell
inequality for a singlet-state pair of entangled qubits. Here we
adapt this proof in the form given by d'Espagnat [8], even if the
proof was first given by Wigner [9].

Let us be more precise about the kind of experiment we are going
to consider. In each system, $S$, of the above ensemble, we take
two immediately consecutive measurements of $M$ for two
independent values, randomly chosen, of the three fixed external
parameter values $a$, $b$, and $c$. We will call each of these
pairs of measurements a \textit{run}. Then, as we have explained
before, we assume the existence of a joint reality behind any
obtainable measurement outcome. Let us consider the number of
these joint realities $(a^{\alpha }b^{\beta }c^{\gamma })$, which
are present \textit{after the first measurement of every run and
before the second measurement}. Let us denote this number by
$N(a^{\alpha }b^{\beta }c^{\gamma })$. We will define
\begin{equation}
N(a^{+}b^{-})\equiv  N(a^{+}b^{-} c^{+})+ N(a^{+}b^{-}c^{-})
\end{equation}
\begin{equation}
N(a^{+}c^{-})\equiv  N(a^{+}b^{+} c^{-}) + N(a^{+}b^{-}c^{-})
\end{equation}
\begin{equation}
N(b^{+}c^{-})\equiv  N(a^{+}b^{+} c^{-}) + N(a^{+}b^{-}c^{-})
\end{equation}
From this, we readily have:
\begin{equation}
N(a^{+}c^{-})\le(N(a^{+}b^{-})+N(b^{+} c^{-})).
\end{equation}

Now, let us consider, for example, the number of \textit{runs},
$N[a^{+}b^{-}]$, where $a^{+}$ is the outcome of the first
measurement and $b^{-}$ the outcome of the second one. (Notice
that we use square brackets for measurement outcomes and standard
brackets for hypothetical realities). Obviously, these
\textit{runs} can only come from the above realities $(b^{-})$
between the first and the second measurement. Furthermore, from
the perfect correlation assumption, they can only come from the
more specific realities $(a^{+}b^{-})$. (The notation $(b^{-})$
and $(a^{+}b^{-})$ should be obvious). Then, given a reality
between both measurements such as $(a^{+}b^{-})$, what is the
probability of obtaining a\textit{ run} like $[a^{+}b^{-}]$? Since
the choice of any one of the three parameters, $a$, $b$, $c$, is a
random choice, this probability is just $1/9$. This means that we
can write
\begin{equation}
N(a^{+}b^{-}) = 9N[a^{+}b^{-}],
\end{equation}
and similarly for $N[a^{+}c^{-}]$ and $N[b^{+}c^{-}]$. Thus,
taking into account Eq. (4), we obtain the temporal Bell
inequality:

\begin{equation}
N[a^{+}c^{-}]\le N[a^{+}b^{-}] + N[b^{+}c^{-}],
\end{equation}
for the observable quantities $N[a^{+}c^{-}]$, $N[a^{+}b^{-}]$ and
$N[b^{+}c^{-}]$.

Notice that in this proof it is essential to define the above
joint reality $(a^{\alpha }b^{\beta}c^{\gamma })$ as the joint
reality which is present before the second measurement of the
\textit{run} and after the first one. In this way, the reality can
only be changed by the second measurement. But this change is
irrelevant to the completion of our proof, since in a \textit{run}
we do not consider a third measurement.

If one prefers to speak in terms of probabilities corresponding to
the numbers in inequality (6), we can write this inequality as
\begin{equation}
P(a^{+},c^{-})\le (P(a^{+},b^{-})+P(b^{+},c^{-})),
\end{equation}
and in a similar way
\begin{equation}
P(a^{-},c^{+})\le (P(a^{-},b^{+})+P(b^{-},c^{+})).
\end{equation}

Or, in terms of the expected value,
\begin{equation}
E(a,b) =
P(a^{+},b^{+})+P(a^{-},b^{-})-P(a^{+},b^{-})-P(a^{-},b^{+}),
\end{equation}
taking into account inequalities (7) and (8), we obtain:
\begin{equation}
E(a,b)+E(b,c)-E(a,c)\le 1.
\end{equation}

At first sight, one might think that inequalities (7), (8), or
(10) are of no interest since, if they were experimentally
violated, this could always be explained by some transmission of
information between the two consecutive measurements of the
\textit{run}. But this is not true since, as we have seen,
inequalities (7), (8), and (10) have been deduced from the joint
realism assumption, using the perfect correlation property,
without any further assumptions. Therefore, we can transmit all
kinds of information we want between both measurements, but if
perfect correlation and joint realism are preserved, as we assume,
inequalities (7), (8), and (10) must remain true.

Now, we could find that the joint realism assumed in the present
paper is a too restrictive postulate in the case of a quantum
system, and, thus, a non convincing postulate for such a system.
In fact, in quantum mechanics, two non commuting observables
cannot be measured at the same time. Furthermore, the orthodox
interpretation of the theory assumes that it is not only that we
cannot jointly measure them, but it asserts that the corresponding
joint reality does not exist. On the other hand, as we will see in
the next Section, quantum mechanics entails the violation of our
Bell inequalities. Then, we can say that the non existence of
joint reality in the case of a qubit is not a question of
interpretation, but a prediction of quantum mechanics which could
be easily tested experimentally, by testing these Bell
inequalities. Obviously, it is to be expected that experience will
agree in this point with quantum mechanics and so that it will
reject the assumed joint reality.

\section{3. Quantum violation of the new Bell inequalities }

Let us assume that our system \textit{S} is a qubit. Then, a
normalized general state, $|\psi>$, can be written as:
\begin{equation}
|\psi> = s |e+> + (1- s^2)^{1/2} e^{i \phi} |e->,
\end{equation}
where $|e+>$ and $|e->$ are the eigenstates of eigenvalues $\pm
1$, respectively, for a given ``direction'' $e$. Since for any
``direction'' $x$ the corresponding eigenstates, $|x+>$ and
$|x->$, are orthogonal unit vectors in a 2-dimensional Hilbert
space, it is straightforward to show that an angle $\alpha_x$ and
a phase $\phi$, always exist such that
\begin{equation}
|x+> =[(1+\cos\alpha_x)/2]^{1/2} |e+> + [(1-\cos\alpha_x)/2]^{1/2}
e^{i\phi} |e->,
\end{equation}
\begin{equation}
|x-> =[(1-\cos\alpha_x)/2]^{1/2} |e+> -[(1+\cos\alpha_x)/2]^{1/2}
e^{i\phi}|e->.
\end{equation}

This means, as it is well-known, that $x$ and $e$ can always be
interpreted as two unit 3-vectors in $\textbf{R}^{3}$,
$\textbf{x}$ and $\textbf{e}$, respectively, which appear in these
equations only through their 3-scalar product $\textbf{x.e} = \cos
\alpha_{x}$.

Hence, when measuring the above dichotomic magnitude \textit{M} for the
three external parameter values, \textit{a}, \textit{b}, and \textit{c}, we
can always say that these measurements have been taken for the corresponding
unit 3-vectors, \textbf{\textit{a}},\textbf{\textit{ b}},
and\textbf{\textit{ c}}.

Let us consider the different probabilities, $P(\textbf{a}^{\pm
},\textbf{b}^{\pm })$, of obtaining $\pm $1 for the two
consecutive measurements of the \textit{runs} where chance has
selected, respectively, the unit 3-vectors $\textbf{a}$ and
$\textbf{b}$. After some basic algebra, we find
\begin{equation}
P(\textbf{a}^{+},\textbf{b}^{-}) = s^{2}(1-\textbf{a.b}) /2,
P(\textbf{a}^{-},\textbf{b}^{+}) = (1-s^{2})(1-\textbf{a.b}/2.
\end{equation}

Thus, according to Eq. (9), we obtain:
\begin{equation}
E(\textbf{a},\textbf{b}) = \textbf{a.b},
\end{equation}
which differs in sign from the similar result for the expected
value in the case of an entangled pair of qubits in the singlet
state. (Obviously, for $E(a,c)$ and $E(c,b)$, we have similar
equations to Eq. (15)).

Notice that the result (15) has the remarkable property of being
independent of the initial state of the particle [10], that is, in
(15), $E(\textbf{a},\textbf{b})$ does not depend on $s$ or $\phi$
appearing in Eq. (11), while$P(\textbf{a}^{\pm },\textbf{b}^{\pm
})$ does depend on $s$. All this means that the version (10) of
our Bell inequalities does not depend on the initial state of the
system $S$, while versions (7) or (8) do.

Bearing in mind Eq. (15) and the similar ones, the Bell inequality
(10) becomes
\begin{equation}
\textbf{a.}(\textbf{b}-\textbf{c})+\textbf{b.c}\le 1,
\end{equation}
which is clearly violated by any two orthogonal unit 3-vectors
$\textbf{b}$ and $\textbf{c}$, if the unit 3-vector $\textbf{a}$
is collinear to $\textbf{b}-\textbf{c}$. In this case, the left
hand side of inequality (16) takes the value $\surd $2.

Similarly one can see that the quantum mechanics of qubit (11)
violates inequalities (7) or (8), but this violation depends on
the initial state of the qubit. For example, if this initial state
is the eigenstate $|a+>$, for the different probabilities
appearing in inequality (7) one finds:
\begin{equation}
P(a^{+},c^{-}) = (1-\textbf{a.c})/2, P(a^{+},b^{-}) =
(1-\textbf{a.b})/2, P(b^{+},c^{-})
=(1+\textbf{a.b})(1-\textbf{b.c})/4.
\end{equation}

Then, for inequality (7) we get:
\begin{equation}
\textbf{b.}(\textbf{a}+\textbf{c})-2\textbf{a.c}+(\textbf{a.b})(\textbf
{b.c})\le 1.
\end{equation}

For $\textbf{a.c} = 0$ and $\textbf{b} =
(\textbf{a}+\textbf{c})/\surd 2$, this inequality is more severely
violated than the above inequality (16), since the left hand side
becomes $\surd {2}+1/2$ for this configuration, instead of the
above $\surd 2$ for inequality (16). This means that inequalities
(7) or (8) are more severely violated than inequality (10), which
is in fact our version of the original inequalities of Leggett and
Garg. Thus, in the macroscopic domain, we expect that our
inequalities (7) or (8) could be more severely violated than
Leggett and Garg inequalities.

\section{4. Some examples}

Once we have seen that the new Bell inequalities (7), (8), and
(10) can be violated by quantum mechanics, we roughly turn to the
question of how this violation could be experimentally produced.
Here, the problem is that we need to perform two successive
measurements on the same system, and not merely in two different
parts of the same system, as in the ordinary space entangled Bell
inequalities. Then, in the quantum case, we must guarantee that
the first of these two measurements is always a first class
measurement, that is, a preparation-like measurement, in order to
preserve the existence of the system and then be able to take the
second measurement. These conditions can be fulfilled, in
principle, in the case where the measured system is a one half
spin particle, whose spin is successively measured in different
directions, with a Stern-Gerlach device. We must then distinguish
two cases, according to whether we want to test inequality (10),
or alternatively one of the two inequalities (7) or (8).

In the first case, in order to obtain the measurement outcomes to
calculate, for example, the expected value $E(a,b)$ in (10), we
must perform two different measurement series, two \textit{run}
series to be more precise (following a similar strategy to the one
stated in [6], where the authors combine different ideal
negative-result setups). One \textit{run} series to obtain the
probabilities $P(a^{+},b^{+})$ and $P(a^{+},b^{-})$ and the other
\textit{run} series to obtain the probabilities
\textit{P(a}$^{-}$\textit{,b}$^{-}$\textit{) }and\textit{
P(a}$^{-}$\textit{,b}$^{+}$\textit{)}. In the first series we only
retain the \textit{a}$^{+}$ outcomes corresponding to a
preparation-like measurement. In this way, the spin particle is
still available for a second measurement in direction
\textit{b}\textbf{.} We will proceed in a similar way for the
second series, where in another large and identical ensemble we
will only retain the \textit{a}$^{-}$ outcomes corresponding to
another preparation-like measurement. In this way, we will be able
to measure the three expected values \textit{E(a,b)},
\textit{E(b,c)} and \textit{E(a,c)}, and thus to test inequality
(10). Notice that, as we have already remarked, in the present
case we do not need to prepare the system in any particular state
before each \textit{run}.

Nevertheless, it could be doubted whether this method, of
performing two different series of measurements, can guarantee the
existence of a ''common probability space'', as argued in Ref.
[11]. Then, to overcome this unfair state, we can consider the
second case: the case corresponding to, let us say, inequality
(7), which is, furthermore, an interesting case in itself. In this
second case, we assume that the $a^{+}$ and $b^{+}$ outcomes in
(7), related to the first measurements of each \textit{run}, refer
to preparation-like measurements. In the present case, the
different probabilities which appear in (7) do depend on the
particle state before each run. Then, for each run, we will
prepare this previous state as an $a^{+}$ eigenstate. This is what
has been assumed in order to deduce the inequality (18) that, as
we have seen, is more severely violated than the corresponding
inequality (16). In all, for each run, we must perform three
successive Stern-Gerlach measurements: first, we must prepare the
$a^{+}$ eigenstate, and then perform two successive measurements
from this eigenstate. In this way we will be able to measure the
three probabilities of inequality (7) and thus to test this
inequality.

Let us emphasize that, as requested in Ref. [11], when testing
inequality (7), there is a ''common probability space'' behind
inequalities (7) or (8), though this is simply forced by joint
reality assumption, which by itself means the common existence of
different settings for the same event of the sample space.

Thus, since we can scarcely doubt that inequality (7), if tested,
would be experimentally violated, we must conclude that joint
realism is ruled out, not by any quantum mechanics interpretation,
but by quantum mechanics itself. As we have already commented,
this would be so, without any concern about locality conditions,
since the proof of the new Bell inequalities, we are considering
here, do not rely on the locality assumption (no locality loophole
can be present here). Furthermore, according to what we have just
explained, the existence of a common probability space would also
be guarantied.

On the other hand, in the case of macroscopic systems, the problem
of the possible measurements which destroy the system is absent.
So, the problem of having more than a single probability space is
also absent, by two different reasons. First of all, because it is
forced by the joint reality assumption. Also, because now, in the
case of inequality (10), differently to what happens in the
microscopic case, we do not need to rely on the strategy of two
different measurement series.

Finally, one could consider a micrometer sized super-conducting
loop, with Josephson junctions [12], to test  our Bell
inequalities in the macroscopic case, as it has been considered by
several authors [7]. But it seems to us that it would be more
interesting to use our temporal Bell inequalities to test realism
in the case of mesoscopic dichotomic random systems whose
randomness could be not so obviously retraced to an enclosed
quantum system.

\section{6.-Conclusions}

In the present paper, we have proved some Bell inequalities for
consecutive measurements under the assumptions of ``joint
realism'', using the perfect correlation property, for any kind of
physical system, macroscopic or microscopic, with a randomly
dichotomic magnitude, i.e., a magnitude which randomly takes two
values. The measurement outcomes are the response of the system to
some different external parameter values, as in the standard Bell
inequalities. These different parameter values play the role of
the different measurement times in the seminal paper of Legget and
Garg [6]. In the paper, these authors deal with realism and NIM
assumptions in the context of macroscopic systems. In the present
paper, we deal jointly with macroscopic or microscopic systems, by
substituting both assumptions for the joint reality assumption,
and by using the perfect correlation property. Contrarily to the
case of NIM assumption, joint reality can be asserted, in
principle, either for microscopic systems or for macroscopic ones.

On the other hand\textbf{,} the perfect correlation property is always
verified by quantum systems. When the physical system is a macroscopic one,
one must verify whether the perfect correlation property is satisfied. One
can expect that this property will be verified in the macroscopic case on
the grounds of the joint reality assumption\textbf{.}

The new assumption of joint reality substitutes NIM assumption
advantageously because that joint reality assumption can be
applied, in principle, to quantum systems too, and also because it
has provided us with new Bell inequalities for consecutive
measurements, which can be expected to be more severely violated
than the temporal ones from Leggett and Garg. Then, notice that,
when trying to prove our Bell inequalities, if joint reality is
assumed, and perfect correlation holds, we do not need to be
concerned with any kind of information which could be propagated
between the two measurements of a \textit{run}.

\acknowledgements This work has been supported by the Generalitat
Valenciana GV05/264.

I am grateful to E.Santos for fruitful discussions.


\begin{thebibliography}{0}

\bibitem{1} BELL J. S., \textit{Physics} \textbf{1} (1964) 195.

\bibitem{2} ClAUSER, J. F., HORNE M. A., SHIMONY A. and HOLT R. H., \textit{Phys.
Rev.} \textit{Lett}. \textbf{23} (1969) 880.

\bibitem{3} PAZ J. P. and MAHLER G., \textit{Phys. Rev. Lett}. \textbf{71 }(1993)
3235.

\bibitem{4} HUELGA S. F., MARSHALL T. W. and SANTOS E., \textit{Phys. Rev. A}
\textbf{54} (1996) 1798.

\bibitem{5} HUELGA S. F., MARSHALL T. W. and SANTOS E., \textit{Europhys. Lett}.
\textbf{38} (1997) 249.

\bibitem{6} LEGGETT A. J. and GARG A., \textit{Phys. Rev. Lett}. 54 (1985) 857.

\bibitem{7} BALLENTINE L. E., \textit{Phys. Rev. Lett.}, \textbf{59} (1987) 1493~;
PERES A., \textit{Phys. Rev. Lett., }\textbf{61} (1988) 2019;
CALARCO T. and ONOFRIO R., \textit{Phys. Lett. A}, \textbf{198}
(1995) 279; \textbf{208} (1995) 40; BENATTI F., GHIRARDI G. C. and
GRASSI R., \textit{Nuovo Cimento B}, \textbf{110 }(1995) 593. See
also the two responses of Leggett and Garg: LEGGETT A. J. and GARG
A., \textit{Phys. Rev. Lett.}\textbf{ 59 }(1987) 1621; \textbf{63}
(1989) 2159.

\bibitem{8} D'ESPAGNAT B., \textit{Scientific American} \textbf{241, }num. 5 (1979)
158.

\bibitem{9} WIGNER E. P., \textit{American Journal of Phys.} \textbf{38 }(1970)
1005.

\bibitem{10} BRUKNER \u {C}., TAYLOR S., CHEUNG S. and VEDRAL V.,
\textit{arXiv.quant}-\textit{ph}/\textit{0402127 v1}.

\bibitem{11} HESS K. and PHILLIP W., \textit{Foundations of
Physics}\textbf{35} (2005) 1749.

\bibitem{12} VAN DER WAL C. \textit{et al}, \textit{Science}, \textbf{290} (2000)
773
\end{thebibliography}
\end{document}